\documentclass[pra,aps,amsmath,amssymb,twocolumn,groupedaddress ,floatfix,hidelinks]{revtex4-2}
\usepackage{graphicx,
            amsmath,
            braket,
            units,
            ragged2e,
            xcolor,
            xspace,
            nicefrac,
            lipsum,
            nameref,
            textcomp,
            urwchancal,
            upgreek,
            isotope,
            hyperref,
            cleveref,
            placeins, 
            scrextend, 
            url}
\usepackage{amstext}
\usepackage{amsfonts}
\usepackage{amsxtra}
\usepackage{bm}
\usepackage{grffile}
\usepackage{soul}
\usepackage{epstopdf}
\usepackage{marvosym} 
\usepackage{tikz}
\usepackage{flushend}
\usepackage{scalerel}

\usepackage[normalem]{ulem}

\usetikzlibrary{svg.path}

\definecolor{nclblue}{RGB}{0,70,127}
\definecolor{nclred}{RGB}{198,12,48}

\definecolor{lancsred}{RGB}{181,18,27}
\definecolor{lancsgrey}{RGB}{85,86,86}
\definecolor{lancsgreen}{RGB}{56,97,74}	
\definecolor{lancsblue}{RGB}{32,84,121}	
\definecolor{lancspalegrey}{RGB}{190,192,194}
\definecolor{lancsorange}{RGB}{253,144,41}

\hypersetup{
    colorlinks=true,
    allcolors = blue
}

\DeclareMathAlphabet{\mathcal}{OMS}{cmsy}{m}{n}

{%
\setlength{\fboxsep}{0pt}%
\setlength{\fboxrule}{1pt}%
}%

\definecolor{orcidlogocol}{HTML}{A6CE39}
\tikzset{
  orcidlogo/.pic={
    \fill[orcidlogocol] svg{M256,128c0,70.7-57.3,128-128,128C57.3,256,0,198.7,0,128C0,57.3,57.3,0,128,0C198.7,0,256,57.3,256,128z};
    \fill[white] svg{M86.3,186.2H70.9V79.1h15.4v48.4V186.2z}
                 svg{M108.9,79.1h41.6c39.6,0,57,28.3,57,53.6c0,27.5-21.5,53.6-56.8,53.6h-41.8V79.1z M124.3,172.4h24.5c34.9,0,42.9-26.5,42.9-39.7c0-21.5-13.7-39.7-43.7-39.7h-23.7V172.4z}
                 svg{M88.7,56.8c0,5.5-4.5,10.1-10.1,10.1c-5.6,0-10.1-4.6-10.1-10.1c0-5.6,4.5-10.1,10.1-10.1C84.2,46.7,88.7,51.3,88.7,56.8z};
  }
}

\newcommand\orcidicon[1]{\href{https://orcid.org/#1}{\mbox{\scalerel*{
\begin{tikzpicture}[yscale=-1,transform shape]
\pic{orcidlogo};
\end{tikzpicture}
}{|}}}}

\newcommand{\rr}{\mathbf{r}}

\newcommand{\Arg}{\text{Arg}}

\begin{document}
\title{Deterministic Nucleation and Dynamics of Infilled Multiply-Charged Vortices in an Immiscible $^{87}\mathrm{Rb}$--$^{41}\mathrm{K}$ Mixture}

\author{R. Doran\,\orcidicon{0000-0002-9467-1264}}
\email{r.doran@lancaster.ac.uk}
\affiliation{Joint Quantum Centre (JQC) Durham--Newcastle, School of Mathematics, Statistics and Physics, Newcastle University, Newcastle upon Tyne, NE1 7RU, United Kingdom}
\affiliation{MARS: Mathematics for AI in Real-world Systems, School of Mathematical Sciences, Lancaster University, Lancaster, LA1 4YF, United Kingdom}

\author{K. E. Wilson\,\orcidicon{0000-0001-5683-8441}} \email{kali.wilson@strath.ac.uk}
\affiliation{Department of Physics, SUPA, University of Strathclyde, Glasgow G4 0NG, United Kingdom}

\date{\today}

\begin{abstract}
We propose a method for controllably generating multiply-charged vortices in immiscible Bose-Einstein condensates. We achieve this by applying a laser stirring technique to a $^{87}\mathrm{Rb}$--$^{41}\mathrm{K}$ mixture, where the vortices generated are infilled by the secondary component. We numerically demonstrate that the charge of the vortex can be tuned reproducibly by varying the stirring parameters, allowing the deterministic generation of stable infilled vortices with high topological charge. We then consider the dynamics of these multiply-charged vortices in a circular trap; in contrast to single-component condensates, we observe long-lived precession of the multiply-charged vortices with a charge dependent frequency and collective breathing modes of the infilling component. For specific large winding numbers, we observe distinct dynamical instabilities leading to vortex dislocation. 
\end{abstract}

\maketitle

\section{Introduction}
\label{section:introduction}
Turbulence is responsible for the transport of energy in fluids, underpinning phenomena that range from atmospheric and ocean dynamics, to biological applications and aerodynamic engineering. As one of the fundamental building blocks of turbulence, it is hardly surprising that vortices have attracted great interest in superfluids formed of ultra-cold atomic Bose Einstein condensates (BECs). Here, unlike in a classical fluid, vortices are topologically protected entities with a fixed core size and a circulation that is quantized in units of Planck's constant divided by the mass of the atomic species. Vortices can form spontaneously in a BEC in a variety of ways, such as through thermal quenches \cite{Weiler2008} similar to the Kibble-Zurek mechanism \cite{Kibble1976,Zurek1985} or by global phase transitions such as the Berezinskii-Kosterlitz-Thouless transition in a 2D system \cite{Hadzibabic2006}. Vortices can also be formed using mechanical methods: forcing the condensate to rotate by stirring with a laser spoon \cite{Madison2000,AboShaeer2001} or spinning the trapping potential \cite{Hodby2001} leads to a regular triangular lattice of vortices, dragging a repulsive (blue-detuned) optical potential through the condensate faster than the critical velocity will nucleate vortices in the wake \cite{Inouye2001,Kwon2015,Kwon2016}, while dragging an attractive (red-detuned) optical potential through the condensate excites standing modes that decay into vortices \cite{Onofrio2000}.

Of critical importance to experimental studies of vortex dynamics in superfluid BECs is the ability to controllably add vortices with a given charge to a system at a given position. This is particularly important in studies of turbulent BECs, where it is necessary to obtain a quasi-steady state of turbulence \cite{Wilson2013}, which rely on statistical averaging and therefore require initial vortex configurations to be reproducible \cite{Barenghi_review_2014}. Vortices were first deterministically created in a BEC by imprinting a relative phase on a two-component BEC \cite{Matthews1999}. Later developments employed a so-called ``spiral'' technique where magnetic push coils dragged the condensate around a repulsive potential, stirring a pinned vortex into the bulk of the condensate before unpinning the vortex by ramping off the laser \cite{Samson2016,Wilson2022}. More recently, the advent of Digital Micromirror Devices (DMDs) that can create arbitrary spatio-temporal potentials in a condensate \cite{Gauthier2016}  have found use in creating more complex dynamical stirring protocols. This has immediate application to the spiral technique, allowing for multiple beams to imprint several vortices at once \cite{Wilson2022,Neely2024}, as well as the deterministic creation of vortex--antivortex pairs \cite{Roati_chopsticks}, and chiral vortex gases \cite{Reeves2022}.

Binary mixtures of BECs have been achieved with an atomic species in different hyperfine states \cite{Myatt1997,Hall1998,Matthews1999,Miesner1999,Maddaloni2000,Delannoy2001,Schweikhard2004,Mertes2007,Anderson2009,Tojo2010}, two isotopes of the same atomic species \cite{vanKempen2002,Papp2008}, and different atomic species \cite{Ferrari2002,Modugno2002,Thalhammer2008,McCarron2011}. The dynamics of the two condensates are coupled by inter-species interactions, and the mixture can be tuned to be either miscible or immiscible \cite{Pu1998}. Mixtures give rise to a wide variety of exotic steady \cite{Pu1998,Ho1996,Timmermans1998,Ao1998,Trippenbach2000,Barankov2002,VanSchaeybroeck2008,Guatam2010,Gordon1998,Kim2002} and dynamic \cite{Li2019,Han2019,He2019, Mithun2021,Wheeler2021,Richaud2023,McMillan2024} solutions. Of particular interest is the immiscible regime, where the atoms of one component will ``infill'' the vortex cores of the other component, since these are density-depleted regions \cite{Matthews1999,Anderson2000}. Depending on the system parameters (for example, the strength of immiscibility between the two components and the ratio of atom numbers), the infilling component can trace the fluid flow \cite{Spielman2025} or the position of the vortex cores \cite{Doran2022_variational}, or it can drive new forms of vortex dynamics \cite{Richaud2020,Richaud2021,Richaud2023}, be responsible for vortex shedding \cite{Sasaki2011,McMillan2024} and Onsager vortex clustering \cite{Han2019}, or stabilize multiply-charged vortices \cite{Patrick2023}. In the latter case, multiply-charged vortices, which would typically be unstable to separation in single component systems \cite{Adhikari2019}, are stabilized due to the immiscibility of the infilling component \cite{Patrick2023}. 

\begin{figure}
    \centering
    \includegraphics{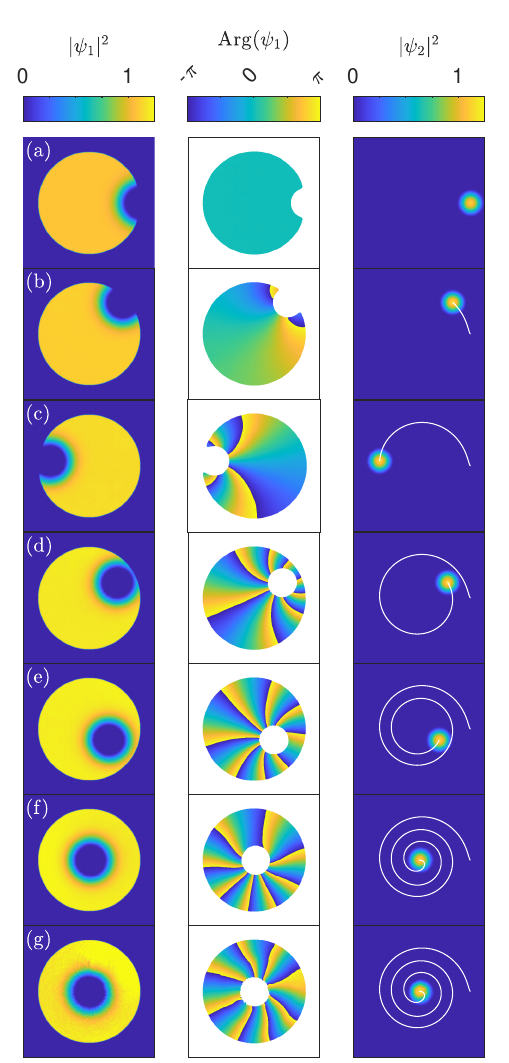}
    \caption{Stills from a simulation of the spiral technique with $N_s=3$ and $T_s=10^{4}$, which results in an infilled vortex with charge $q=8$ at the origin. Here the beam width $w=32$, the trap radius $R=100$, and $N_2=1000$. The first column shows the density of component 1, $|\psi_1|^2$; the second column shows the phase of component 1, $\Arg\left(\psi_1\right)$; the third column shows the density of component 2, $|\psi_2|^2$. For clarity, we only plot the phase of component 1 where $|\psi_1|^2>0.01$, and the trajectory of the spiral beam is added in white to the third column as a guide. The rows correspond to different times, (a) at $t=0$, (b) at $t=2\times10^3$, (c) at $t=4\times10^3$, (d) at $t=6\times10^3$, (e) at $t=8\times10^3$, (f) at $t=10^4=T_s$, and (g) at $t=1.2\times10^4$ (in dimensionless variables). A movie of this simulation is available in the Supplemental Material \cite{Supplemental}.}
    \label{fig:stirring_Ns3_Tr10}
\end{figure}

In this paper, we extend the stirring technique described in Refs.~\cite{Samson2016,Wilson2022} to an immiscible $^{87}\mathrm{Rb}$--$^{41}\mathrm{K}$ mixture, see Fig.~\ref{fig:stirring_Ns3_Tr10}. We numerically demonstrate that the spiraling technique may be applied to this mixture, creating a vortex at a determined position in the bulk of the $^{87}\mathrm{Rb}$ component whose core is filled by the $^{41}\mathrm{K}$ component. We explore how the spiral parameters determine the charge of the vortex, and show that the resulting multiply-charged infilled vortex is stable. We then consider the dynamics of a multiply-charged infilled vortex away from the center of the trap, and consider the precession dynamics of this vortex.

The remainder of this paper is structured as follows: In Section~\ref{section:governing_equations} we present the equations of motion for the immiscible mixture. In Section~\ref{section:vortex_nucleation} we describe the stirring technique for an immiscible mixture and explore the parameter space for vortex nucleation. In Section~\ref{sec:dynamics} we consider the dynamics of a massive infilled vortex within a circular trap, observing both its precession rate and the breathing modes of the infilling component. Conclusions and future work are given in Section~\ref{section:conclusions}.

\section{Equation of Motion for a $^{87}\mathrm{Rb}$--$^{41}\mathrm{K}$ Mixture}
\label{section:governing_equations}
In the zero-temperature limit, a mixture of two weakly interacting dilute gases of bosons are described by the macroscopic wavefunctions $\Psi_1$ and $\Psi_2$. The energy of the system can be written as $E^{(\mathrm{3D})}=\int d^3\rr \ \mathcal{E}^{(\mathrm{3D})}\left(\Psi_1,\Psi_2\right)$, where
\begin{eqnarray}
    \mathcal{E}^{(\mathrm{3D})} &=& \frac{\hbar^2}{2m_1}\left|\nabla \Psi_1\right|^2 + \frac{\hbar^2}{2m_2}\left|\nabla\Psi_2\right|^2 \nonumber \\
    &+& \mathcal{V}_1\left(\rr,t\right)\left|\Psi_1\right|^2 + \mathcal{V}_2\left(\rr,t\right)\left|\Psi_2\right|^2 \nonumber \\
    &+& \frac{1}{2} U_{11} \left| \Psi_1\right|^4 +  U_{12} \left|\Psi_1\right|^2 \left| \Psi_2 \right|^2 + \frac{1}{2}U_{22} \left|\Psi_2\right|^4\,, \qquad  
    \label{eqn:3D_energy}
\end{eqnarray}
with $m_j$ the mass and $\mathcal{V}_j$ the external trapping potential acting on the $j$--th component. The \emph{intra}-species interaction strengths are given by
\begin{equation}
    U_{jj}=\frac{4\pi\hbar^2a_{jj}}{m_j}\,,
    \label{eqn:full_intra}
\end{equation}
for $j\in\{1,2\}$, while the \emph{inter}-species interaction strength is given by
\begin{equation}
    U_{12}=\frac{2\pi\hbar^2a_{12}}{m_r}\,,
\end{equation}
where $a_{j,k}$ parameterizes the $s$-wave scattering length between atoms in the $j$--th and $k$--th component, and the reduced mass is given by $m_r = m_1 m_2/\left(m_1 + m_2\right)$. In order to consider a quasi-2D system, we impose a harmonic trapping on the system in the $z$ direction so that
\begin{equation}
    \mathcal{V}_j\left(\rr, t\right) = V_j\left(x,y,t\right) + \frac{1}{2} m_j \omega_j^2 z^2\,,
\end{equation}
where $V_j$ is a (time-dependent) potential in the $xy$ plane, and the $\omega_j$ are sufficiently strong that the static harmonic trapping potential will prevent excitations along the $z$ axis \cite{Rooney2011}. The fully three dimensional wavefunction of each component may then be written as 
\begin{equation}
    \Psi_j\left(\rr,t\right) = \left(\ell_j\sqrt{\pi}\right)^{-1/2} \psi_j\left(x,y,t\right)\exp\left(-z^2/2\ell_j^2\right)\,,
    \label{eqn:3D_wavefunction}
\end{equation}
where $\ell_j=\sqrt{\hbar/m_j\omega_j}$ is the harmonic oscillator length in the $z$ direction. We can then find the corresponding energy of the quasi-2D system by substituting the 3D ansatz, Eq.~\eqref{eqn:3D_wavefunction}, into the energy function, Eq.~\eqref{eqn:3D_energy}, and integrating over $z$ to obtain $E^{(\mathrm{2D})}=\int\,d^2\rr\,\mathcal{E}^{(\mathrm{2D})}\left(\psi_1,\psi_2\right)$, where 
\begin{eqnarray}
    \mathcal{E}^{(2D)} &=& \frac{\hbar^2}{2m_1}\left|\nabla_\perp \psi_1\right|^2 + \frac{\hbar^2}{2m_2}\left|\nabla_\perp \psi_2\right|^2 \nonumber \\ &+& V_1\left(x,y,t\right)\left|\psi_1\right|^2 + V_2\left(x,y,t \right)\left|\psi_2\right|^2 \nonumber \\ &+& \frac{1}{2} u_{11} \left| \psi_1\right|^4 + u_{12} \left|\psi_1\right|^2 \left| \psi_2 \right|^2 + \frac{1}{2} u_{22} \left|\psi_2\right|^4\,. \qquad 
    \label{eqn:2D_energy}
\end{eqnarray}
Eq.~\eqref{eqn:2D_energy} contains the effective 2D \emph{intra}-species interaction strengths $u_{jj}=U_{jj}/\sqrt{2\pi}\ell_j$, and the effective 2D \emph{inter}-species interaction strength $u_{12}=U_{12}/\left(\sqrt{\pi}\sqrt{\ell_1^2+\ell_2^2}\right)$. We have written $\nabla_\perp=\left(\partial/\partial x, \partial/\partial y\right)$ to emphasize that this is now an effective 2D system. In order to recover the coupled Gross-Pitaevskii Equation (GPE) from the 2D energy, we calculate 
\begin{equation}
    i\hbar \frac{\partial \psi_j}{\partial t} = \frac{\delta}{\delta \psi_j^*}E^{(2D)}\,, 
\end{equation}
which yields
\begin{subequations}
    \begin{align}
        i \hbar \frac{\partial \psi_1}{\partial t} &= \left[ -\frac{\hbar^2}{2m_1} \nabla_\perp^2 + V_1 +  u_{11}|\psi_1|^2 + u_{12}|\psi_2|^2 \right]\psi_1\,,  \\
        i \hbar \frac{\partial \psi_2}{\partial t} &= \left[-\frac{\hbar^2}{2m_2}\nabla_\perp^2 + V_2 + u_{12}|\psi_1|^2 + u_{22}|\psi_2|^2 \right]\psi_2\,. 
    \end{align}
\end{subequations}

Without loss of generality, we take component 1 to be the main component and rescale the system using the natural units of this component. We introduce the healing length, $\xi_1 = \hbar/\sqrt{m_1 u_{11} n_{1,0}}$, where $n_{1,0}$ is the density of component 1 in the absence of a stirring beam, and a characteristic time scale $\tau_1 = \hbar / \left(u_{11}n_{1,0}\right)$. We can then introduce dimensionless variables (indicated by a prime) for the wavefunctions $\psi_j = \sqrt{n_{1,0}}\psi^\prime_j$, time $t=\tau_1t^\prime$, spatial derivatives $\nabla_\perp=\xi_1^{-1}\nabla_\perp^\prime$ and trapping potentials $V_j = n_{1,0} u_{11} V_j^\prime$, so that the coupled GPE becomes
\begin{subequations}
    \begin{align}
    i\frac{\partial \psi_1^\prime}{\partial t^\prime} &= \left[ \quad -\frac{1}{2}\nabla_\perp^{\prime 2} + V^\prime_1 + \ \ |\psi^\prime_1|^2 + \alpha |\psi^\prime_2|^2\right]\psi^\prime_1\,, \\
    i\frac{\partial \psi^\prime_2}{\partial t^\prime} &= \left[-\frac{1}{2} \overline{m} \nabla_\perp^{\prime 2} + V^\prime_2 + \alpha |\psi^\prime_1|^2 + \beta |\psi^\prime_2|^2 \right] \psi^\prime_2\,.
    \end{align}
    \label{eqn:main_GPE}
\end{subequations}
We have introduced the mass ratio $\overline{m}=m_1/m_2$, as well as two dimensionless parameters, the ratio of inter-species and intra-species interaction strengths
\begin{equation}
    \alpha = \frac{u_{12}}{u_{11}} = \frac{1}{\sqrt{2}} \frac{a_{12}}{a_{11}} \frac{\left(1+\overline{m}\right)}{\sqrt{1+\overline{m}\overline{\omega}}}\,,
\end{equation}
and the ratio of intra-species interaction strengths
\begin{equation}
    \beta = \frac{u_{22}}{u_{11}} = \frac{a_{22}}{a_{11}} \sqrt{\frac{\overline{m}}{\overline{\omega}}}\,,
\end{equation}
where $\overline{\omega}=\omega_1/\omega_2$. In the remainder of this paper we will not write the primes. 

If component 1 is $^{87}\mathrm{Rb}$ and component 2 is $^{41}\mathrm{K}$, then the mass ratio is $\overline{m}=2.1217$. We take $\overline{\omega}=0.7345$, which corresponds to a harmonic trapping potential that is imposed in the $z$ direction by a $1064\mathrm{nm}$ wavelength laser. The dimensionless parameters $\alpha$ and $\beta$ can be tuned using Feshbach resonances (for Rb-K see, for example \cite{Ferlaino_Feshbach_KRb, Thalhammer2008, Burchianti_Mixture_KRb}), here we take $\beta=1.0999$ (corresponding to $a_{11}=100.44a_0$ and $a_{22}=65.0a_0$ where $a_0$ is the Bohr radius) and choose $\alpha=2.1242$ (corresponding to $a_{12}=154.6a_0$) so that the mixture is in the immiscible regime \cite{Trippenbach2000}. Numerically we evolve the coupled GPE, using \textsc{xmds2} \cite{XMDS2}, with an adaptive Runge-Kutta method in an $L_x \times L_y$ doubly periodic grid with $N_x\times N_y$ points; throughout this work we take $L_x=L_y=256$ and $N_x=N_y=512$.

\section{Vortex Nucleation}
\label{section:vortex_nucleation}

\subsection{Stirring Technique}
\label{subsec:stirring_process}
In a single-component system, the vortex generating technique works by spiraling a repulsive optical potential (blue-detuned laser beam) from the edge of the condensate to the desired vortex position. Initially the beam induces a superflow that creates a superfluid circulation pinned to the centre of the repulsive potential. The potential can then be moved to a desired position in the bulk of the condensate, translating the pinned superflow to this location. The beam is then ramped off after a hold time, and the vortex core is formed. See Refs.~\cite{Samson2016,Wilson2022} for full details. In a single component system, multiply-charged vortices are energetically unfavourable \cite{Patrick2023}, meaning that the multiply-charged superflow decays into a cluster of singly charged vortices when the pinning potential is removed \cite{Wilson2022}. 

We apply the stirring technique to an immiscible mixture of $^{87}\mathrm{Rb}$ and $^{41}\mathrm{K}$ atoms. In our system, we expect that the number of $^{87}\mathrm{Rb}$ atoms is much greater than the number of $^{41}\mathrm{K}$ atoms (i.e., $N_1>N_2$), and so our objective is to controllably generate a vortex in $^{87}\mathrm{Rb}$ (component 1) whose core is infilled by $^{41}\mathrm{K}$ (component 2); we achieve this by utilizing a stirring potential that is repulsive in component 1 and attractive in component 2. The spiraling technique is very similar to the one-component system: the optical potential induces a superflow in component 1, this creates a superfluid circulation that is locked to the centre of the potential. The potential simultaneously acts as a confining potential for component 2, which then results in an infilled vortex once the potential is ramped off. In contrast to the single-component system, we expect that for sufficient infilling atom numbers, the system will support multiply-charged infilled vortices. Experimentally this can be implemented in a $^{87}\mathrm{Rb}$--$^{41}\mathrm{K}$ mixture by using an optical potential with a wavelength of approximately $775~\mathrm{nm}$ such that the polarisability of the beam is opposite but equal, providing a blue-detuned (repulsive) beam in $^{87}\mathrm{Rb}$ (component 1) and a red-detuned (attractive) beam in $^{41}\mathrm{K}$ (component 2).  The spiraling of the beam with respect to the atoms can be achieved either by using magnetic push coils \cite{Wilson2022}, or by using a digital micromirror device (DMD) \cite{Gauthier2016}; the latter of which provides scope to impose multiple spiraling beams \cite{Neely2024}. 

We decompose the trapping potentials $V_1$ and $V_2$ in Eqn.~\eqref{eqn:main_GPE} as 
\begin{equation}
    V_1(x,y,t) = V_\mathrm{HW}(x,y) + V_\mathrm{Obst}(x,y,t)\,,
\end{equation}
and 
\begin{equation}
    V_2(x,y,t) = - V_\mathrm{Obst}(x,y,t)\,, 
\end{equation}
where $V_\mathrm{HW}$ is a stationary hard-walled trapping potential, defined as $V_\mathrm{HW}(x,y) =0$ when $x^2+y^2<R^2$, and $V_\mathrm{HW}(x,y) =10$ otherwise. The time-dependent stirring potential is given by
\begin{equation}
    V_\mathrm{Stir}(x,y,t) = V_0 \exp\left\{ -\frac{\left[x-x_s\left(t\right)\right]^2 + \left[y-y_s\left(t\right)\right]^2}{w^2}  \right\}\,,
    \label{eqn:stirring_potential}
\end{equation}
where $\left(x_s,y_s\right)$ is the center of the spiraling beam. The spiral trajectory of the stirring beam can be written in polar coordinates as 
\begin{equation}
    r_s = R\sqrt{1-\frac{t}{T_s}}\,, \qquad \theta_s = 2\pi N_s \left(\frac{t}{T_s}\right)^2\,,
\end{equation}
where $R$ is the radius of the hard-walled trap, $N_s$ is a parameter that determines the number of complete spirals that the beam will complete, and $T_s$ is the time taken for the beam to spiral to the center of the trap \cite{Wilson2022}. This means that the position of the stirring beam in Eqn.~\eqref{eqn:stirring_potential} is given as $\left(x_s,y_s\right)=\left(r_s\cos\theta_s,r_s\sin\theta_s\right)$. The radius of the zero-density region in the majority component that is due to the stirring potential can be approximated from the Thomas-Fermi profile as $2w\sqrt{\ln\left(V_0\right)}$. 

An example of the stirring technique can be found in Fig.~\ref{fig:stirring_Ns3_Tr10}. As we will show in the next section, the charge of the pinned superfluid circulation depends on the width of the beam, $w$, and the speed at which the beam spirals into the system (determined by $N_s$ and $T_s$). As has previously been discussed, the exact trajectory of the spiraling beam is unimportant in controllably generating circulation (and ultimately vortices) in a BEC, so long as the initial motion is roughly tangent to the hard-wall trap \cite{Wilson2022}. Of far greater importance is that $N_s$ and $T_s$ are chosen so that the speed of the beam is low enough to prevent other excitations in the system. Once such excitation is the shedding of vortex dipole pairs in component 1, which can occur either because the repulsive potential travels through the bulk of the condensate above a critical velocity for vortex nucleation \cite{Kwon2015}, or because vortices de-pin from the stirring potential. A second excitation is the formation of a dark-soliton that occurs when the beam is entirely contained within the trapping potential and the fluid merges [see Fig.~\ref{fig:stirring_Ns3_Tr10} row (d)], which can happen if the beam is moving too quickly \cite{Wilson2022}.

\subsection{Vortex Winding Number}

\begin{figure}
    \centering  \includegraphics{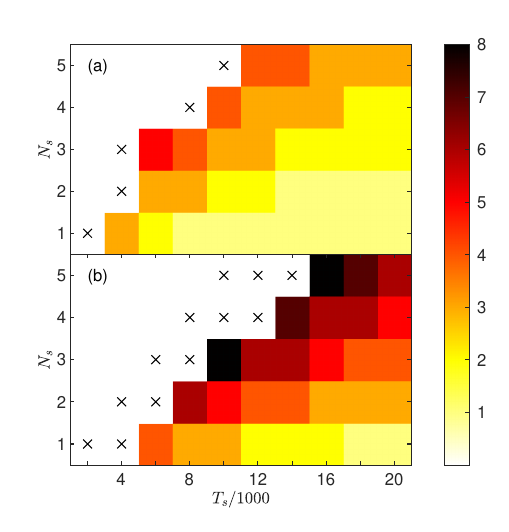}
    \caption{The resulting phase winding pinned at the origin as a function of $T_R$ and $N_s$ for (a) $w=16$ and (b) $w=32$. Black crosses (x) indicate values of $T_s$ and $N_s$ where the stirrer moved faster than the critical velocity for vortex nucleation and additional vortices were shed.}
    \label{fig:vortex_charge}
\end{figure}

We have previously stated that the charge of the infilled vortex created by the stirring technique depends on the beam width, $w$, the number of spirals of the beam, $N_s$, and the time during which the beam spirals, $T_s$. We explore this parameter space, setting $V_0=\exp(1)$ and $N_2=1000$. We choose the atom number of the majority component so that the background density is approximately unitary in the absence of a stirring potential, that is $N_1=\pi R^2$. For our system, $R=100$ (in dimensionless variables) so $N_1\approx 3.1\times10^4$, and therefore $N_2/N_1 \approx 0.03$. This is much greater than the critical ratio that Ref.~\cite{Patrick2023} predicted to support multiply-charged vortices. In order to initialize the system, we evolve the GPE, Eqn.~\ref{eqn:main_GPE}, in imaginary time with the stirring potential at the edge of the stationary trap [see Fig.~\ref{fig:stirring_Ns3_Tr10}, row (a)], before evolving the system in real time without additional damping.

The results of these simulations can be found in Fig.~\ref{fig:vortex_charge}. For the beam widths that we consider [$w=16$ in panel (a) and $w=32$ in panel (b)], the majority of the pinned superflows that we create are multiply-charged. In order to generate the highest winding numbers, the stirring beam needs to be moving relatively quickly (corresponding to a high number of revolutions, $N_s$, and a short stirring time, $T_s$), however not so quickly that excitations are formed away from the beam. The wider beam provides higher circulation for comparable $\left(T_s,N_s\right)$, which is to be expected given that the beam acts as a pinning potential, and wider pinning potentials can support higher winding numbers \cite{Doran2024_disorder}. The wider beams also have a lower critical velocity for vortex nucleation \cite{Frisch1992}, meaning that a larger value of $T_s$ is required for a given $N_s$ to avoid de-localized excitations. For each of the values of $N_s$ we see that the resulting circulation has a weak dependence on the stirring time, $T_s$. For large $T_s$ and low $N_s$, corresponding to a very low stirring speed, we see that only one quanta of circulation is pinned to the stirring potential, which is the lowest value that we observed for our parameters. 
 
\begin{figure}
    \centering
    \includegraphics{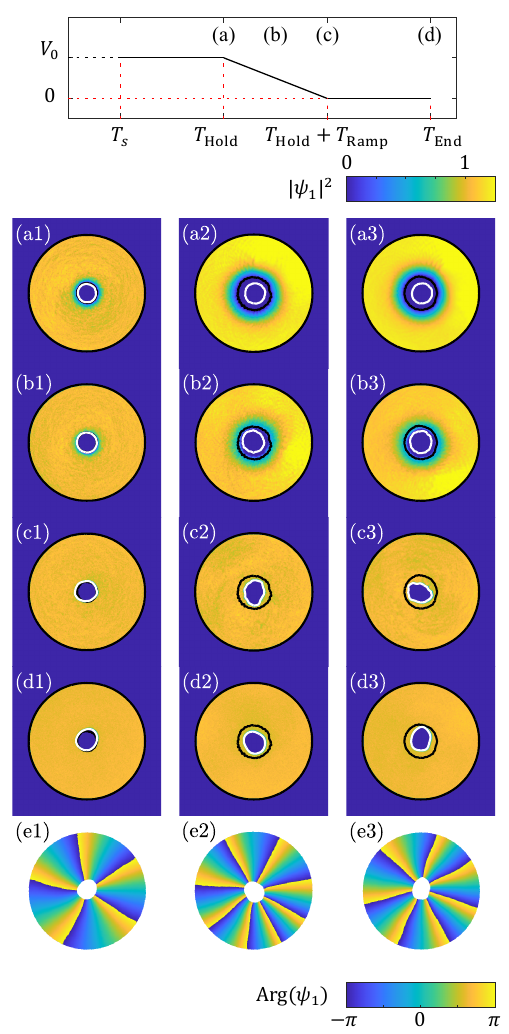}
    \caption{Snapshots of the vortex created at the origin as the obstacle potential is ramped off. The top panel shows a schematic of the ramp-off protocol. Rows (a)-(d) show the density of the main component, $|\psi_1|^2$, row (e) shows the phase of the main component, $\Arg\left(\psi_1\right)$. Snapshots in a given row are taken at the same time, the time of each row corresponds to text in the top panel. As before, we only plot the phase of component 1 where $|\psi_1|^2>0.01$. As a guide to the eye, the black density contour $|\psi_1\left(T_\mathrm{Hold}\right)|^2=0.01$ is plotted for each set of stirring parameters. The white contour displays $|\psi_2(t)|^2=0.5$. The columns correspond to different stirring parameters:  
    Column (1), $w=16$, $N_s=3$, $T_s=6\times10^3$ creates a charge $q=5$ vortex; Column (2), $w=32$, $N_s=3$, $T_s=10^4$ creates a charge $q=8$ vortex; Column (3), $w=32$, $N_s=4$, $T_s=1.4\times10^4$ creates a charge $q=7$ vortex. }
    \label{fig:ramp_off}
\end{figure}

\subsection{Ramp-Off Stability at the Origin}
\label{section:stability}
Having shown that it is possible to deterministically create multiply-charged pinned superfluid circulation, we now consider the stability of the superflow as the pinning potential is slowly ramped off. In a single component system, we would expect that generated vortices with a winding number $|q|>1$ would split into $|q|$ singly-charged vortices as the potential is ramped off \cite{Wilson2022}. However, as we are in an immiscible binary system, we expect that for some choice of filling atom number $N_2$ and interaction parameters $\alpha$ and $\beta$, multiply-charged vortices can be stabilized \cite{Patrick2023}.  

The procedure for removing $V_\mathrm{Stir}$ is as follows: we allow the system to evolve as described in Sec.~\ref{subsec:stirring_process} until $t=T_s$, at which point the stirring potential is stationary at the origin. We then hold the stirring potential fixed until $T_\mathrm{Hold}=T_s+2000$, this is to allow the sound waves generated by the movement of the stirring potential to equilibrate. The strength of the stirring potential is then ramped down according to 
\begin{equation}
    V_0(t) = \exp(1)\left(1-\frac{t}{T_\mathrm{Ramp}}\right)\,,
\end{equation}
when $t\leq T_\mathrm{Ramp}$, and $V_0=0$ otherwise. We take $T_\mathrm{Ramp}=2000$, and allow the system to evolve until $T_\mathrm{End}=2.5\times10^4$ and observe the multiply-charged infilled vortex.

The results of the ramping-off process can be seen in Fig.~\ref{fig:ramp_off}. We note that Ref.~\cite{Patrick2023} performed a linear stability analysis for an infilled vortex with charge $q=2$ at the centre of a circular hard-walled trap, and found that the dynamic stability of the multiply-charged infilled vortex depended on the number of infilling atoms. For our choice of $N_2$ and $R$, the highest winding numbers produced by our stirring protocol are $q=5$ (when $w=16$, $N_s=3$ and $T_s=6\times10^3$) and $q=8$ (when $w=32$, $N_s=3$ and $T_s=10^4$); the stability of vortices with these winding numbers is presented in Fig.~\ref{fig:ramp_off} columns 1 and 2, respectively. We also examine the dynamic stability of a $q=7$ vortex (that may be generated when $w=32$, $N_s=4$ and $T_s=1.4\times10^4$) in Fig.~\ref{fig:ramp_off} column 3, as we observe that this winding number is dynamically unstable when the vortex is off-centre (see Sec.~\ref{sec:dynamics}). We find that the multiply-charged infilled vortices are stable against splitting for the lifetime of the simulation. In the cases of the wider stirring-beam, $w=32$ in columns 2 and 3, as the stirring beam is ramped off the vortex core relaxes to a size that is energetically favorable \cite{Doran2022_variational}. Since the vortex is initially located at the origin, we would expect that the vortex remains at the origin as the image vortex is located at infinity. The slightly non-circular nature of the vortex core may be due to the way in which the vortex is dynamically stirred into the condensate and the small amount of sound that remains in the system, but may also be influenced by shape oscillations of the infilling component (see Sec.~\ref{subsec:breathing_modes}). We also observe that the branch-cuts in the phase, corresponding to the vortex location in the point vortex limit, are not exactly co-located, but are contained within the enlarged vortex core created by the infilling component.

\begin{figure}
    \centering
    \includegraphics{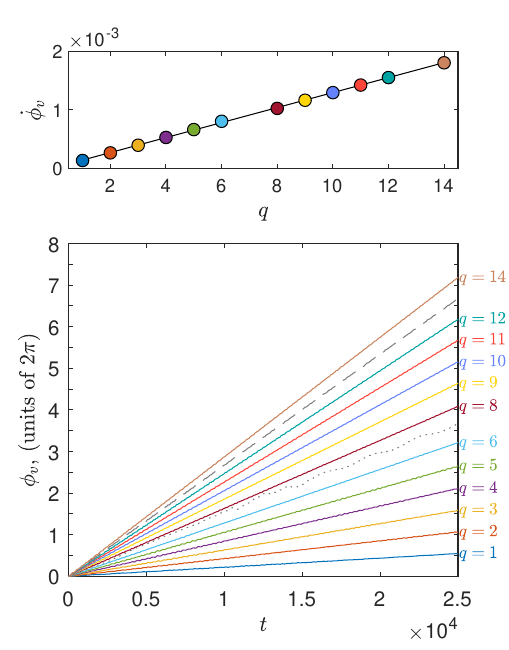}
    \caption{Main panel: The precession of the vortex, $\phi_v$, as a function of time. The charge of the vortex, $q$, is indicated; the infilling atom number, $N_2$, and the interaction parameters, $\alpha$ and $\beta$, are all fixed. Upper panel: The extracted precession frequency of the vortex, $\dot{\phi}_v$, for each of the vortex charges; the black line shows the least squares fit $\dot{\phi}_v=Aq+B$ with $A=1.279\times10^{-4}$ and $B=1.705\times10^{-5}$.}
    \label{fig:precession_plot}
\end{figure}


\section{Multiply-Charged Vortices}
\label{sec:dynamics}

\subsection{Dynamics}
We now turn our attention to the dynamics of an infilled vortex with charge $q\geq1$ that is displaced from the centre of the trap. It has previously been shown that infilled vortices display novel dynamics that are unseen in their core-less counterparts \cite{He2019,Richaud2023,McMillan2024}, however to the best of our knowledge, this is the first study of the dynamics of stable multiply-charged vortices in a binary BEC. We begin by positioning a vortex with charge $q$ at $\left(0,R/2\right)$ and allowing the system to evolve. We initialise the vortex at $R/2$ so that it will be moved by its image vortex but it is not so close to the edge of the trap that the infilling component will leave the vortex core. Throughout this section we will focus on $1\leq q \leq 14$, as we found that $q\geq 15$ can't be stabilized for our choice of $\alpha$, $\beta$ and infilling atom number. We consider the dynamic instability of $q=7$ and $q=13$ in Sec.~\ref{subsec:dynamic_instability}.

To track the infilled multiply-charged vortex, we calculate the centre of the mass of the infilling component, given by
\begin{equation}
    \rr_\mathrm{C.o.M} = \frac{1}{N_2}\int \, d^2\rr \ \rr |\psi_2(\rr)|^2\,,
    \label{eqn:centre_of_mass}
\end{equation}
which gives us the vortex position $\left(r_v,\phi_v\right)$ in polar coordinates. The precession of the vortex is shown in Fig.~\ref{fig:precession_plot}. The vortex precesses about the trap at a near constant angular velocity, moved by the velocity field of its image vortex. As predicted by the massive point vortex model \cite{Richaud2020,Richaud2021}, we observe oscillatory behavior in the radial component of the velocity due to the inertial effect of the infilling component. This can be seen in the small waves in the main panel of Fig.~\ref{fig:precession_plot}. Having extracted the centre of the infilling component, we fit the precession to $\phi_v = \dot{\phi}_v t + \phi_{v,0}$, where $\dot{\phi}_v$ is the precession frequency of the vortex and $\phi_{v,0}\approx 0$ is a constant \cite{Freilich2010}. For each value of $q$, we fit this linear model with a coefficient of determination that is greater than $0.9999$. The extracted value of the precession frequency as a function of the vortex charge can be found in the upper panel of Fig.~\ref{fig:precession_plot}. The precession frequency increases linearly with the vortex charge, in agreement with the massive point vortex model prediction that 
\begin{equation}
    \dot{\phi}_v = \eta q  \left[1-\sqrt{ 1  - \frac{2}{\eta\left(R^2 - r_0^2\right)}}\, \right] \,,
    \label{eqn:mpvm_precession_frequency}
\end{equation}
for an initial vortex displacement $r_0$ and with $\eta = N_1 \overline{m}/\left(N_2 R^2\right)$ (see Appendix~\ref{appendix_massivepv} for further details). For our parameters, $\eta\approx 6.67\times 10^{-3}$ and so Eq.~\eqref{eqn:mpvm_precession_frequency} gives $\dot{\phi}_v\approx1.35\times10^{-4} q$, which is in good agreement with our least squares fit of $A=1.279\times10^{-4}$.
\begin{figure}
    \centering
    \includegraphics{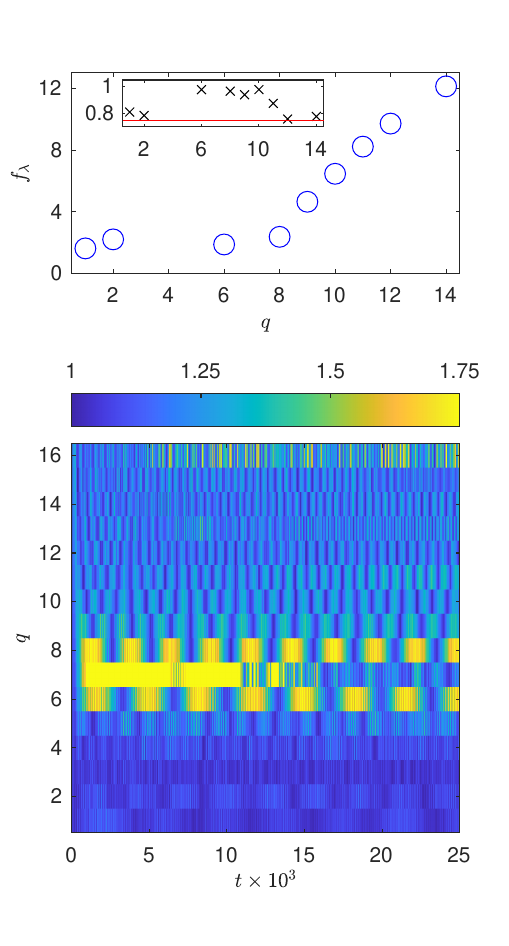}
    \caption{Main panel: The aspect ratio of the infilling component, given by Eq.~\eqref{eqn:aspect_ratio}, as a function of time. The charge of the vortex, $q$ is indicated on the vertical axis, the colourbar represents $\lambda$. Upper panel: The extracted frequency of the aspect ratio, $f_\lambda$, as a function of vortex charge, found by fitting Eq.~\eqref{eqn:fourier}; the inset shows the coefficient of determination for each value of $q$. Note we only present extracted frequencies where the coefficient of determination is greater than 0.75, indicated by the solid red line.}
    \label{fig:breathing}
\end{figure}

\subsection{Breathing Modes}
\label{subsec:breathing_modes}
In the immiscible limit, the infilling component feels an effective trapping potential that is due to the vortex core. As such, the infilling component is subject to small amplitude oscillations, similar to the breathing or collective modes that occur in a single-component system in the presence of an external trap \cite{Pethick_and_Smith}, or to the shape oscillations observed in classical droplets \cite{Staat2017}. In the two component case, the breathing modes of the infilling component have an additional degree of complexity, in that the effective trapping potential, formed by the vortex core, can alter shape and radiate small density waves (in the form of sound waves) every time the vortex accelerates.

In order to study the shape oscillations that occur in the component that fills the vortex core, we compare the number density of the infilling component to a Gaussian profile given by
\begin{equation}
    n_2 = A \exp\left\{-\frac{1}{2}\left[\left(\frac{\tilde{x}}{\sigma_x}\right)^2 + \left(\frac{\tilde{y}}{\sigma_y}\right)^2\right]\right\}\,, 
\end{equation}
where $\tilde{x}(t)=\left[x-x_c(t)\right]\cos\varphi - \left[y-y_c(t)\right]\sin\varphi$ and $\tilde{y}(t)=\left[x-x_c(t)\right]\sin\varphi+\left[y-y_c(t)\right]\cos \varphi$, with $\left(x_c,y_c\right)$ the center of mass coordinates of the infilling component given by Eq.~\eqref{eqn:centre_of_mass}. This choice of coordinate system allows the Gaussian to rotate from the $x$ axes by an angle $\varphi$, which occurs as the vortex precesses about the hard-walled trap. Of particular interest to the breathing modes of the system is the aspect ratio of the infilling component, which is approximated by the ratio of the maximum and minimum Gaussian widths,
\begin{equation}
    \lambda = \frac{\max\left(\sigma_x\,,\,\sigma_y\right)}{\min\left(\sigma_x\,,\,\sigma_y\right)}\,.
    \label{eqn:aspect_ratio}
\end{equation}
A consequence of this is that $\lambda\geq1$, with $\lambda=1$ corresponding to the infilling component having a (roughly) circular density profile. The results of this fit can be found in the main panel of Eq.~\ref{fig:breathing}. Generally we see that the frequency with which the aspect ratio oscillates increases with the charge of the vortex, this is particularly apparent in the vortices where $8\leq q \leq 12$.  In the case where $q=7$, the infilling component becomes extremely elongated before vortices are dislocated; we discuss this case further in the next section.

To further explore the connection between the vortex charge and the oscillations of the aspect ratio of the condensate trapped in the vortex core, we extract the frequency of the oscillations by fitting a Fourier profile of the form
\begin{equation}
    \lambda(t) = a_0 + a_1\cos\left(f_\lambda \frac{t}{1000}\right) + b_1\sin\left(f_\lambda \frac{t}{1000} \right)\\,
    \label{eqn:fourier}
\end{equation}
where $f_\lambda$ is the dimensionless frequency of the aspect ratio. We note that before fitting Eq.~\eqref{eqn:fourier}, we first smooth the data with a Savitzky-Golay filter, and we exclude data before $t=5000$ to reduce the noise from the initial acceleration of the vortex. We observe that the behaviour of the frequency of the aspect ratio can be divided into two distinct regimes, with a turnover around the $q=7$ instability. For vortices with charge $1\leq q \leq 5$, there is relatively little oscillation in the aspect ratio, which may suggest that the kinetic energy imparted by the vortex motion is insufficient to excite significant breathing modes in the infilling component. We caveat this claim by reporting that the signal to noise ratio is particularly poor for these values of $q$, so it is difficult to extract a precise scaling here. For vortices with charge $8\leq q \leq 12$, there is a clear linear scaling that increases with $q$. This is indicative that the increase in momenta of the main component about the vortex core is driving stronger surface modes on the infilling component. Future work in this area may wish to consider the effect of different atom numbers in the infilling component, in particular in determining whether the charge at which there is a turnover between the two scaling regimes can be reduced or increased by lowering or raising the atom number, respectively. Additionally, more complex profiles could be fitted to the infilling component, as the Gaussian profile is a first order approximation that is ill suited to the irregular shapes seen in some of the simulations (see Fig.~\ref{fig:instability}).

\subsection{Instability}
\label{subsec:dynamic_instability}

\begin{figure}
    \centering
    \includegraphics{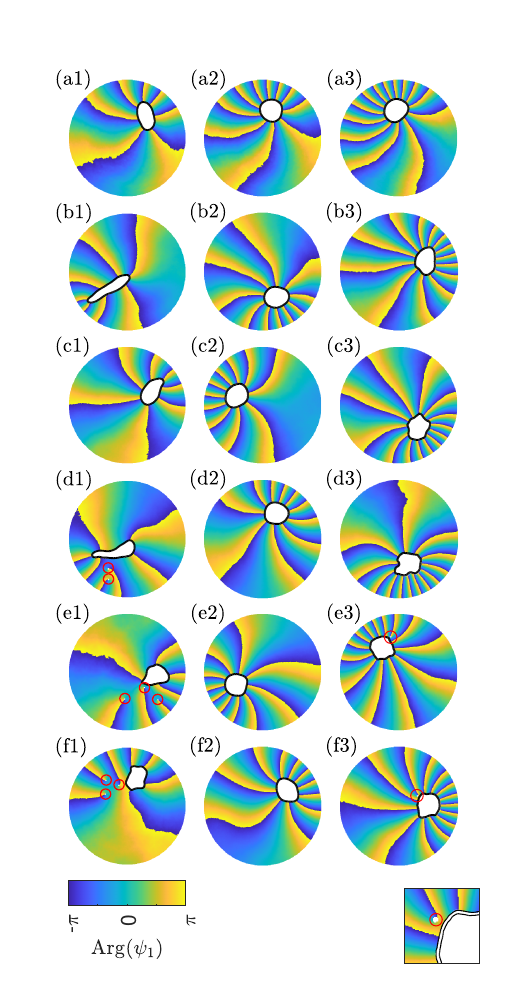}
    \caption{Snapshots of the phase of the off-center vortex as it precesses around the trap. Columns correspond to different vortex charges: Column 1, q=7 (unstable); Column 2, q=10 (stable); Column 3, $q=13$ (unstable). As a guide to the eye, dislocated vortices are highlighted with a red circle, and the density contours $|\psi_2|^2=\{0.1,0.5\}$ are plotted in black. Rows show the temporal evolution of the phase at (a) $t=10^3$, (b) $t=4\times10^3$, (c) $t=7\times10^3$, (d) $t=1.05\times10^4$, (e) $t=1.25\times10^4$, and (f) $t=1.5\times10^4$. The bottom right inset is a zoomed in view of panel (f3). A movie of this simulation is available in the Supplemental Material \cite{Supplemental}.}
    \label{fig:instability}
\end{figure}

We now consider the case of  multiply-charged vortices--- stabilized with an infilling component--- with $q=7$ and $q=13$, which we observed to be dynamically unstable. The manifestation of these instabilities is shown in Fig.~\ref{fig:instability}. We note that, although both of the instabilities result in a dislocation of one or more vortices from the infilled vortex, the process by which this occurs is fundamentally different. 

In the case where $q=7$, the multiply-charged infilled vortex initially precesses about the trap although it is distinctly anisotropic in shape, as is shown in the main panel of Fig.~\ref{fig:breathing}. The branch-cuts that would indicate the position of singly-charged vortices in the point vortex limit are irregularly distributed about the infilled vortex core; in fact, as can be seen in Fig.~\ref{fig:instability}~(b1) and (d1), vortices cluster at the extreme points of the multiply-charged core. Eventually, the energy associated with having an elongated vortex core \cite{Barankov2002,VanSchaeybroeck2008} (the analogue of a large surface area in 3D) overtakes the energetic cost to break the vortex up, and singly-charged vortices are dislocated from the core. As a result, the remaining multiply-charged infilled vortex is able to adopt a more circular shape [Fig.~\ref{fig:breathing}, also Fig.~\ref{fig:instability}~(e1) and (f1)], and this remaining multiply-charged vortex precesses about the trap. This precession is shown in the dotted curve in Fig.~\ref{fig:precession_plot}; while the rate agrees qualitatively with Eq.~\eqref{eqn:mpvm_precession_frequency}, for $q=7$ it is clear that the assumption of uniform precession no longer applies. The singly charged vortices that are created can be ``re-captured'' by the original multiply-charged vortex, but their dynamics are typical of like-sign vortices in that they co-rotate and their distance from the original vortex becomes large. We note that typically less that 0.1\% of the infilling atoms are transferred from the original multiply-charged infilled vortex to the vortices that are shed.

The process by which the $q=13$ vortex breaks up is remarkably different. Throughout the precession of this vortex (dashed line in Fig.~\ref{fig:precession_plot}), the infilling component remains roughly circular (Fig.~\ref{fig:breathing}), and the distribution of branch-cuts in the phase is relatively uniform across the contour, as can be seen in Fig.~\ref{fig:instability}~(a3) and (b3). Rather than the vortex core elongating, the edge of the core develops a node corresponding to a branch-cut in the phase, Fig.~\ref{fig:instability}~(c3) and (d3). This node appears to travel around the edge of the vortex core, although its presence is not enough to drastically alter the aspect ratio of the infilling component. Eventually, a singly-charged vortex is dislocated at the location of this node, Fig.~\ref{fig:instability}~(e3). The dislocated vortex precesses about the original (now $q = 12$) vortex, although, unlike in the $q=7$ case, the distance between the dislocated vortex and the original vortex is relatively small, Fig.~\ref{fig:instability}~(f3), presumably because the azimuthal velocity of the now $q=12$ vortex is dominating the dynamics of the dislocated vortex. It should be emphasized that the $q=14$ vortex is stable for the lifetime of our simulations, and so we believe that there is a sufficient infilling component to stabilize the $q=13$ vortex \cite{Patrick2023}; the appearance of a node on the edge of the vortex core is suggestive that the instability of the $q=13$ vortex may be due to a packing problem within the core. 

Further classification of the instabilities seen by multiply-charged infilled vortices is an avenue for future work. In particular, it would be interesting to investigate these instabilities as the radius of the hard-walled trap, $R$, varies, thereby altering the influence of the image vortex, as well as investigating the role of the infilling atom number, $N_2$, where larger atom numbers should stabilize higher charges.

\section{Conclusions}
\label{section:conclusions}

In this work we have demonstrated a deterministic method for the creation and control of multiply-charged vortices with filled cores in an immiscible binary BEC, and we have explored the rich dynamical behavior that emerges when multiply-charged infilled vortices freely evolve. These results are of particular importance to the future of experiments that will focus on turbulence in two-component superfluids, as they provide a method for engineering reproducible vortex configurations \cite{Barenghi_review_2014}. 

We have extended a stirring technique, originally developed for single-component condensates \cite{Wilson2022}, to an immiscible $^{87}\mathrm{Rb}$--$^{41}\mathrm{K}$ mixture by leveraging specific wavelengths in which an optical potential is repulsive for one component but attractive for the other. This modification enables control over both the vortex core and the infilling component, controllably nucleating multiply-charged infilled vortices that are stabilized by the infilling component. By systematically exploring the stirring parameter space, we have shown that the vortex charge can be tuned reproducibly over a large range. Crucially, we numerically demonstrate that the resulting multiply-charged infilled vortex is stable in the absence of a pinning potential. Similar to previous demonstrations in single component BECs \cite{Roati_chopsticks, Reeves2022}, this method may be implemented by state-of-the-art digital micromirror devices, enabling unprecedented control over initial conditions in two-component superfluid experiments featuring arbitrary configurations of multiply-charged infilled vortices.

Motivated by the stability of these multiply-charged infilled vortices, we have also investigated the dynamics of vortices when they are placed off-centre in a circular trap. These vortices exhibit long-lived precession that is driven by their image. We have shown that the precession frequency increases linearly with the charge of the vortex, in excellent agreement with the massive point-vortex model \cite{Richaud2020,Richaud2021}. We have also highlighted the role of the dynamics of the infilling component on the vortices. In particular, we have identified breathing modes of the infilling component that are driven by the charge of the vortex. For particular vortex charges, we observed distinct dynamical instabilities that led to vortex dislocation. For charge $q=7$, this instability presented as a strongly anisotropic deformation of the vortex core, while for charge $q=13$, this instability appeared as a singly-charged vortex propagating along the edge of the infilled vortex core. The distinct nature of these instabilities suggest that there is a subtle link between vortex charge, infilling atom number, and confinement that should be explored further.

Experimentally, the ability to deterministically generate stable, high-charge infilled vortices provides a new route to studying quantum turbulence in two-component superfluids, where massive vortices are expected to modify inverse energy cascades \cite{Reeves2013} and clustering phenomena \cite{Wilson2013,Han2019}. Arrays of multiply-charged vortices could also be used to explore Tkachenko modes, vortex lattice melting, or vortex-vortex collisions that are inaccessible to single-component systems. Our work establishes infilled multiply-charged vortices as a versatile and experimentally accessible platform for exploring topological excitations and non-equilibrium dynamics in two-component superfluids.

\acknowledgements 
We thank Andrew Baggaley, Nick Parker and Tapio Simula for interesting discussions about this work.  Numerical simulations made use of the Rocket HPC facility at Newcastle University. This work was supported by Research England under the Expanding Excellence in England (E3) funding stream, which was awarded to MARS: Mathematics for AI in Real-world Systems in the School of Mathematical Sciences at Lancaster University. RD was supported by the UK Engineering and Physical Sciences Research Council, Grant No. EP/X028518/1, and KEW acknowledges support from the Royal Society through a University Research Fellowship (Grant No. URF$\backslash$R1$\backslash$201134).

\appendix

\section{Point Vortex Prediction}
\label{appendix_massivepv}
It is possible to use the massive point vortex model derived in Ref.~\cite{Richaud2020} to predict the precession rate of an infilled vortex trapped in a circular boundary \cite{Richaud2021}. Here, we develop the result in Ref.~\cite{Richaud2021} to be able to describe a multiply-charged infilled vortex. 

We consider the case of a single vortex with charge $q$ that is contained in a fluid with background density $n$ within a circular disk of radius $R$. The Lagrangian for this system is given by
\begin{eqnarray}
    \mathcal{L} &=& \frac{1}{2} M_v \left| \boldsymbol{\dot{r}}_v\right|^2   + 
     \pi n \hbar q \left(\boldsymbol{\dot{r}}_v \times \boldsymbol{r}_v\right)\cdot\boldsymbol{\hat{z}} \nonumber \\
     & \ & \qquad \qquad \ - \ 
     \frac{\pi n \hbar^2 q^2}{m_1} \ln \left(1-\frac{\left|\boldsymbol{r}_v\right|^2}{R^2}\right)\,, 
\end{eqnarray}
where the first term accounts for the massive nature of the vortex (where $M_v = N_2 m_2$ is the mass of the infilling atoms), the second term is the usual vortex velocity term, and the third term accounts for the trapping potential \cite{Richaud2021}, which scales with the square of charge. We recast this system in the characteristic units of the first component: length $\xi_1=\hbar/\sqrt{m_1 u_{11} n_{1,0}}$, time $\tau_1 = \hbar/u_{11} n_{1,0}$, and energy $n_{1,0}u_{11}$. The background density of the fluid is given by the number of atoms, $N_1$, divided by the area of the trap. As a result, the Lagrangian is written in dimensionless variables as
\begin{eqnarray}
    \mathcal{L}' &=& \frac{1}{2} \frac{N_2}{\overline{m}} \left| \boldsymbol{\dot{r}}_v^\prime \right|^2 + q \frac{N_1}{R^{\prime 2}} \left(\boldsymbol{\dot{r}}_v^\prime \times \boldsymbol{r}_v^\prime\right)\cdot \boldsymbol{\hat{z}} \nonumber \\
    & \ & \qquad \qquad \ - \ q^2 \frac{N_1}{R^{\prime 2}} \ln \left( 1 - \frac{\left| \boldsymbol{r}_v^\prime\right|^2}{R^{\prime 2}}\right)\,,  
\end{eqnarray}
where $R=\xi_1 R^\prime$. For the remainder of the appendix we will work in these dimensionless variables without writing the primes.

In polar coordinates, $\left(r_v,\phi_v\right)$, which describe the position of the vortex, the Lagrangian is 
\begin{eqnarray}
    \mathcal{L} &=& \frac{1}{2}\frac{N_2}{\overline{m}} \left(\dot{r}^2_v + r^2_v \dot{\phi}_v^2\right) \  - \  q \frac{N_1}{R^2} r_v^2 \dot{\phi}_v \nonumber \\  
    & \ & \qquad \qquad \  - \ q^2 \frac{N_1}{R^2} \ln\left(\frac{R^2 - r_v^2}{R^2}\right)\,. 
\end{eqnarray}
We are now able to use the Euler-Lagrange equations to write down equations of motion for the azimuthal and radial components of the vortex motion. The equation for the azimuthal motion is 
\begin{equation}
    r_v^2 \left(\frac{N_2}{\overline{m}} \dot{\phi}_v - q \frac{N_1}{R^2}\right) = \mathrm{L}_v\,,
\end{equation}
where $\mathrm{L}_v$ is the angular momentum of the infilled vortex, which is a constant. The radial equation of motion is 
\begin{equation}
    \ddot{r}_v = r_v \dot{\phi}_v^2 - 2\eta q r_v \dot{\phi}_v  +  \frac{2\eta q^2 r_v}{R^2 - r_v^2}\,, 
\end{equation}
where $\eta = N_1 \overline{m}/\left(N_2 R^2\right)$. By re-writing this equation as
\begin{equation}
    \ddot{r}_v = r_v \left[\left(\dot{\phi}_v - q \eta \right)^2 - q^2 \eta^2 \right]  +  \frac{2 \eta q^2 r_v}{R^2 - r_v^2}\,,
    \label{eqn:radial_eqn_motion}
\end{equation}
 we are able to remove the azimuthal dependency from the radial equation, so that 
\begin{equation}
    \ddot{r}_v = r_v \left[ \left(\frac{\overline{m}}{N_2}\right)^2 \frac{\mathrm{L}_v^2}{r_v^4} - q^2 \eta^2 \right]  + \frac{2\eta q^2  r_v}{R^2 - r_v^2}\,. 
    \label{eqn:final_radial_mpvm}
\end{equation}
For a given initial vortex displacement $r_0$, the angular momentum of the infilled vortex, $\mathrm{L}_v$ is fixed. As is argued in Ref.~\cite{Richaud2021}, when $m_2 N_2 / m_1 N_1$ is small (as is the case in our system), the vortex motion approximates uniform precession with a small effect from the mass of the vortex; we see that this is the case in our system by the slight deviation from monotonic behavior in the main panel of Fig.~\ref{fig:precession_plot}. To the lowest order approximation, we then have $\ddot{r}_v\approx 0$ and we can rearrange Eq.~\eqref{eqn:radial_eqn_motion} to find  
\begin{equation}
    \dot{\phi}_v = \eta q   \pm  \sqrt{ \eta^2 q^2  - \frac{2 \eta q^2 }{R^2 - r_0^2}}\,.
    \label{eqn:precessio_freq_pm}
\end{equation}
The negative root of Eq.~\eqref{eqn:precessio_freq_pm} corresponds to the physical, image-driven precession \cite{Richaud2021}, and so we write
\begin{equation*}
    \dot{\phi}_v = \eta q \left[1-\sqrt{1-\frac{2}{\eta\left(R^2-r_0^2\right)}}\right]\,,
\end{equation*}
which is Eq.~\eqref{eqn:mpvm_precession_frequency} in the main text. 
We note that $\lim_{\eta\to\infty} \left[\eta\left(1-\sqrt{1-2 x/\eta}\right)\right]=x$, and so in the limit as $N_2\to 0$ (corresponding to $\eta\to\infty$) we recover the mass-less point vortex limit, $\dot{\phi}_v=q/\left(R^2-r_0^2\right)$. The precession frequency is therefore exactly linear in the vortex charge within this approximation.

\end{document}